\documentclass[prd,preprint,groupedaddress,nofootinbib,showpacs,preprintnumbers,11pt]{revtex4-2}

\pdfoutput=1

\usepackage{epsfig}
\usepackage{hyperref}
\usepackage{amssymb}
\usepackage{amsbsy}
\usepackage{amsmath}
\usepackage{url}

\newcommand{\rf}[1]{(\ref{#1})}
\newcommand{\beq}{\begin{equation}}
\newcommand{\eeq}{\end{equation}}
\newcommand{\be}{\begin{equation}}
\newcommand{\ee}{\end{equation}}
\newcommand{\bea}{\begin{eqnarray}}
\newcommand{\eea}{\end{eqnarray}}
\newcommand{\eq}[1]{Eq.~(\ref{#1})}
\newcommand{\non}{\nonumber \\*}
\newcommand{\ie}{{i.e.}\ }
\newcommand{\bv}{\begin{verbatim}}
\newcommand{\ev}{\end{verbatim}}

\newcommand{\vp}{\varphi}

\newcommand{\e}{\,\mbox{e}}
\renewcommand{\d}{{\rm d}}
\renewcommand{\i}{{\rm i}}

\newcommand{\bz}{{\bar z}}
\newcommand{\p}{\partial}
\newcommand{\bp}{\bar\partial}
\newcommand{\q}{\mbox{}}
\newcommand{\qq}{q}
\newcommand{\hg}{{\hat g}}

\newcommand{\eps}{\varepsilon}
\newcommand{\om}{\omega}

\newcommand{\LA}{\left\langle}
\newcommand{\RA}{\right\rangle}

\def\fun#1#2{\lower3.6pt\vbox{\baselineskip0pt\lineskip.9pt
\ialign{$\mathsurround=0pt#1\hfil##\hfil$\crcr#2\crcr\sim\crcr}}}

\begin{document}

\preprint{}

\title{Exact solution of higher-derivative conformal theory and minimal models}

\author
{Yuri Makeenko}
\vspace*{2mm}
\affiliation{NRC ``Kurchatov Institute''\/-- ITEP,  Moscow\\
\vspace*{1mm}
{\email: makeenko@itep.ru} }

\begin{abstract}

I investigate the two-dimensional four-derivative conformal theory
that emerges from the Nambu-Goto string after the path-integration over all fields
but the metric tensor.
Using the method of singular products which accounts for tremendous cancellations
in perturbation theory, I show the (intelligent) one-loop approximation to give 
an exact solution. It is conveniently described through the minimal models  where
the central charge $c$ in the Kac spectrum depends on the parameters of
the four-derivative action. The relation is nonlinear so the domain of physical
parameters is mapped onto $c<1$ thus bypassing the KPZ barrier of the Liouville action.

\end{abstract}


\maketitle

\newpage

 \section{Introduction}
 
 My original motivation for the research that has resulted in the present Letter was
 the $d=1$ barrier for the existence of bosonic strings first found by
 Knizhnik-Polyakov-Zamolodchikov (KPZ)~\cite{KPZ} for the Polyakov string in 
 the light-cone gauge and then reproduced by David-Distler-Kawai (DDK) \cite{Dav88,DK89}
 in the 
 conformal gauge.
 After quite some efforts I have understood that the only possibility to bypass this 
 no-go theorem is if the Nambu-Goto and Polyakov strings are told apart by higher-derivative
 terms in the emergent action for an independent metric tensor $g_{ab}$.  
 Integrated out
 in the path integral all fields but $g_{ab}$, we arrive in $d$ target-space dimensions
 at the four-derivative emergent action 
\be
{\cal S}[g]=\frac {1}{16 \pi b_0^2} \int \sqrt{g} \left[ - R \frac 1\Delta R +2\mu_0^2
 +\eps  R \left(R+G g^{ab}\,
\partial_a \frac 1\Delta R\;\partial_b \frac 1\Delta R \right)
\right], \quad
 b^2_0=\frac 6{26-d}.~~~
\label{inva}
\ee
Here $R$ is the scalar curvature associated with $g_{ab}$, $\Delta$ is the two-dimensional Laplacian
and $\eps\to0$ is proportional to an ultraviolet cutoff in the path integral. 
For the Polyakov string the action \rf{inva} with $G=0$ 
is derivable from the DeWitt-Seeley expansion of the heat kernel. 
The ${\cal O}(1)$ terms in the conformal gauge become~\cite{Pol81}  
the celebrated Liouville action
while the $R^2$ term  is familiar from the studies~\cite{KN93,Ich95,KSW} 
of the $R^2$ two-dimensional gravity.
The emergence of the second four-derivative term with $G\neq 0$ on the right-hand side of \eq{inva} is specific 
to the Nambu-Goto string~\cite{Mak21}. All other terms with four derivatives can be reduced 
to these two modulo total derivatives. 

The four-derivative terms  in the action~\rf{inva} are suppressed for
smooth metrics as  $\eps R$. However, typical metrics which are essential in the
path integral over  $g_{ab}$ are not smooth and have $R\sim\eps^{-1}$.
This is immediately seen from the action $\rf{inva}$
where the quartic derivative regularizes quantum  divergences with $\eps$ 
playing the role of an {ultraviolet worldsheet cutoff}.  However, $\eps$ is simultaneously a coupling  of
self-interaction of $\vp$ so  uncertainties like $\eps \times \eps^{-1}$ appear in the
perturbation theory.
These uncertainties look like anomalies in quantum field theory  (QFT) and surprisingly affect 
the behavior of strings at large distances.

It worth noting that both for the Polyakov and Nambu-Goto strings the effective action
involves an infinite set
of yet higher-derivative terms, proportional to higher powers of $\eps$. One may expect however
the universality, \ie an independence of the physical results on the precise form of 
higher-derivative terms in the action. For the six-derivative term of the form $\eps^2 R\Delta R$
the universality was demonstrated by explicit computations~\cite{Mak21} and 
proved in~\cite{Mak23b} by another technique. Leaving aside the question about the universality,
I consider in this Letter the four-derivative action~\rf{inva} as such.

It is convenient to fix in the action~\rf{inva} the conformal gauge 
\be
g_{ab}=\hg_{ab}\e^\vp,
\label{confog}
\ee
where  $\hat g_{ab}$ is the background (or fiducial) metric tensor and $\vp$ is the dynamical
 variable often called the Liouville field.
 Naively, one writes for the action \rf{inva} in the gauge \rf{confog}
 \be
{\cal S}^{({\rm min})}[\vp]=\frac 1{16 \pi b_0^2} \int \sqrt{\hg} \left[\hg^{ab} \p_a \vp \p_b \vp  +2\mu_0^2  
+\eps  \e^{-\vp} \hat \Delta \vp
\left(\hat \Delta \vp- G \hat g^{ab}\,\partial_a \vp\partial_b \vp \right) \right] ,~~
\label{invamin}
\ee
where $\hat \Delta$ denotes the Laplacian for the metrics $\hg_{ab}$. However, this is true only 
in the flat background whose curvature $\hat R=0$. As is well-known, under the decomposition~\rf{confog}
\be
\sqrt g R= \sqrt{\hat g} \left( \q\hat R - \hat \Delta \vp \right).
\label{Rshift}
\ee
This results in the appearance of the additional term
$\propto\vp \hat R$ in the Liouville action and analogous terms for the four-derivative
 action~\rf{inva}, resulting in a non-minimal interaction of $\vp$ with background gravity.
 These additional terms are crucial for the Weyl invariance of the four-derivative action.
 
 The  energy-momentum tensor for the minimal action~\rf{invamin} is conserved thanks to the 
 classical equation of motion but not traceless owing to the presence of the dimensionful
 parameters $\mu_0^2$ and $\eps$. 
 The ``improved'' energy-momentum tensor~\cite{DJ95} that
involves additional terms, coming from the non-minimal interaction with background gravity,
is conserved and {\em traceless}\/ thanks to the 
 classical equation of motion for $\vp$ in spite of dimensionful $\mu_0^2$ and $\eps$.
 This is a general property because in the conformal gauge \rf{confog}
we have
\be
T^a_a\equiv\hat g^{ab} \frac{\delta {\cal S}[g]}{\delta \hat g^{ab} }= -
\frac{\delta {\cal S}[g]}{\delta \vp },
\label{tra}
\ee
where the left-hand side of \eq{tra} represents the trace of the 
``improved'' energy-momentum tensor
while the right-hand side represents the classical equation of motion for $\vp$.
The four-derivative action \rf{invamin}
 is thus conformal invariant for flat backgrounds.
  I shall present  explicit formulas for the energy-momentum tensor in the next section. 
 
 The computation of the central charge of $\vp$ at the one-loop order in $b_0^2$
 was performed for the four-derivative action~\rf{inva} by two different methods in~\cite{Mak22}
 and confirmed in~\cite{Mak22c}  by the third one. 
 For $G=0$ there is no difference from the  usual result for the Liouville action. However,
 the additional contribution
 \be
 \delta c^{(\vp)} =6G +{\cal O}(b_0^2)
 \label{cphi}
 \ee
emerges if $G\neq 0$. For the path-integral formulation there is also
a   
 logarithmic divergence which cancels by the one in the string susceptibility and 
is removed in both cases by the normal ordering. 
The addition~\rf{cphi} changes the string susceptibility
for closed surfaces of genus $h$ as~\cite{Mak22,Mak22c} 
\be
\gamma_{\rm str}=
(h-1)\left( \frac {1}{b_0^2}  -\frac76-G +{\cal O}(b_0^2)\right) +2,
\label{gstrfin}
\ee
showing for $G\neq0$ a deviation from the one-loop result~\cite{Zam82,CKT86,KK86}
in the Liouville theory. The conclusion is that the $G$-term is observable quantumly
although is negligible for smooth classical metrics.

The goal of this Letter is to go beyond the one-loop approximation in 
Eqs.~\rf{cphi} and \rf{gstrfin} and to establish 
the connection of the four-derivative conformal theory~\rf{invamin} with
the Belavin-Polyakov-Zamolodchikov (BPZ)~\cite{BPZ} approach to the minimal models.
The main result will be a shift of the KPZ barrier in the Liouville theory.

 \section{``Improved'' energy-momentum tensor}
 
The central role in conformal field theory (CFT) is played by the energy-momentum tensor.
It can be derived by varying the action with respect to $\hg_{ab}$. For the minimal
action~\rf{invamin} we obtain for the $zz$-component of $T_{ab}$ in a flat background
\be
-4 b_0^2 T_{zz}^{({\rm min})}=(\p \vp)^2 -2\eps \p \vp \p \Delta \vp 
-G \eps (\p \vp)^2 \Delta \vp+ 4G\eps\p\vp \p(\e^{-\vp}\p \vp \bp \vp)  
 +  G\q \eps\partial(\p \vp \Delta\vp),
  \label{Tzzmin}
\ee
where $\Delta=4\e^{-\vp}\p  \bp$.
We have $\bp T_{zz}^{({\rm min})}\neq0$ because $T_{ab}^{({\rm min})}$ is not traceless.
Adding the contribution from the non-minimal interaction of $\vp$ with background gravity, 
we find~\cite{Mak22}
\be
T_{zz}= T_{zz}^{({\rm min})}
+\frac1{4b_0^2}\Big[ 2\q\p^2 (\vp - \eps\Delta \vp)  
+ 4G\q\eps \p^2(\e^{-\vp}\p \vp \bp \vp) -
  G \q\eps \frac 1{\bp}\p ^2 (\bp \vp \Delta\vp)\Big],
  \label{Tzz} 
\ee
where the nonlocality of the last term is inherited from the nonlocality of 
the covariant action~\rf{inva}. We now have $\bp T_{zz}=0$ thanks to the classical
equation of motion, explicitly demonstrating the tracelessness of $T_{ab}$ as stated
in the Introduction.

This is however not the whole story because of the renormalization of the parameters 
of the action. Following DDK, I adopt the renormalization of $b_0^2\to b^2$ as well as 
that of \eq{Rshift}
\be
\sqrt g R= \sqrt{\hat g} \left( q\hat R - \hat \Delta \vp \right),
\label{qRshift}
\ee
where $q=1+{\cal O}(b_0^2)$. The origin of the renormalization
is a nontrivial self-interaction of $\vp$ in the four-derivative action~\rf{invamin} as well as
the interaction of $\vp$ with Pauli-Villars' regulators~\cite{Dia89,AM17c,Mak23a}. 
The change $b_0^2\to b^2$ and
$1\to q$ can be viewed as a result of path-integrating over the regulator fields and 
was computed~\cite{Mak22} at one loop by standard diagrams of QFT.
In the CFT technique $b^2$ and $q$ are determined from consistency: the conformal weight
of $\e^\vp$ equals 1 and the total central charge equals 0. The one-loop result
in CFT  remarkably
agrees~\cite{Mak22} with that in QFT.

The renormalization~\rf{qRshift} does not affect $T_{zz}^{({\rm min})}$ but 
the contribution from nonminimal interaction of $\vp$ with background gravity is 
multiplied by $q$. Also $b^2$ substitutes $b_0^2$ in Eqs.~\rf{Tzzmin}, \rf{Tzz}:
\be
 T_{zz}= T_{zz}^{({\rm min})}+\frac1{4b^2}\Big[ 2q\p^2 (\vp - \eps\Delta \vp) + 4G q\eps \p^2(\e^{-\vp}\p \vp \bp \vp)  -
  G q\eps \frac 1{\bp}\p ^2 (\bp \vp \Delta\vp)\Big].
  \label{qTzz} 
\ee
It is the expression for $T_{zz}$ to be used in the CFT technique.

\section{The method of singular products\label{s:dim}}

For a general higher-derivative action  $S[\vp]$ which
is not quadratic in $\vp$ and whose energy-momentum tensor is also not quadratic,  
one sees tremendous cancellations when computing the operator products
 $T_{zz}(z)\e^{\vp(0)}$ or  $T_{zz}(z)T_{zz}(0)$. To account for the cancellation,
 it is convenient to rewrite the generator of 
the conformal transformation as
\be
\hat \delta_\xi \equiv \int_{C_1}\frac{\d z}{2\pi \i} \xi (z) T_{zz} (z)
\stackrel{{\rm w.s.}}
=\int_{D_1} \d^2 z\left(\q \xi'(z) \frac {\delta}{\delta \vp(z)}
 +\xi (z)\p \vp(z)\frac {\delta }{\delta \vp(z)}\right),
 \label{hatdel}
 \ee
where the domain $D_1$ includes the singularities of $\xi(z)$ leaving outside possible singularities of the string of
functions $X(\om_i)$ on which $\hat \delta_\xi $ acts and $C_1$ bounds $D_1$. 
The second equality in \rf{hatdel} is understood in the weak sense, \ie under path integrals.
In proving the equivalence  of the two forms 
we have integrated the total derivative 
\be
\bp T_{zz} =-\pi \q \p \frac {\delta S}{\delta \vp} + \pi \p \vp\frac {\delta S}{\delta \vp}
\ee
and used the (quantum) equation of motion
\be
\frac {\delta S}{\delta \vp} \stackrel{{\rm w.s.}}=\frac {\delta}{\delta \vp}.
\label{26}
\ee
Actually, the form of $\hat \delta_\xi$ on the right-hand side of \eq{hatdel} is primary.
Its advantage over the standard
one on the left is that it accounts for the cancellations  in the quantum case, 
while there are subtleties associated with singular products. 

Averaging over the Pauli-Villars regulators, we arrive  as already discussed at the emergent
action~\rf{invamin} with $b_0^2$ substituted by $b^2$ 
and the energy-momentum tensor \rf{qTzz}. 
The generator~\rf{hatdel} is then substituted by
\be
\hat \delta_\xi = \int_{D_1} \d^2 z \left(\qq \xi' (z)\frac {\delta}{\delta \vp(z)}
 +\xi(z)\p \vp(z)\frac {\delta }{\delta \vp(z)}\right).
 \label{XXq}
\ee

It is easy to reproduce
\be
\hat \delta_\xi \e^{\alpha\vp(\om)} \stackrel{{\rm w.s.}}=
(q \alpha-b^2 \alpha^2)\xi'(\om)\e^{\alpha\vp(\om)} +\xi(\om)\p\e^{\alpha\vp(\om)}
\label{29}
\ee
 for the quadratic action
by \rf{XXq} via the singular products listed in Appendix~\ref{appA}:
\bea
\hat \delta_\xi \e^{\alpha\vp(\om)}   &=&
q \alpha\xi'(\om)  \e^{\alpha\vp(\om)}  +\int_{D_1}\!\d^2 z \, \alpha\xi(z) 
 \p\vp(z)\e^{\alpha\vp(\om)}  \delta^{(2)}(z-\om) \nonumber \\ &\stackrel{{\rm w.s.}}=&
q\alpha \xi'(\om)  \e^{\alpha\vp(\om)} +\!\int_{D_1}\!\d^2 z \, \alpha^2\xi(z) 
\big\langle  \p\vp(z) \vp(\om)\big\rangle \delta^{(2)}(z-\om)  \e^{\alpha\vp(\om)}
 +\alpha \xi(\om)   \p\vp(\om)\e^{\alpha\vp(\om)}.  \non &&
\label{32}
 \eea
 The most interesting is the second term on the right-hand side, where the singular product equals
\be 
\int_{D_1}\!\d^2 z \, \xi(z) 
\big\langle  \p\vp(z) \vp(\om)\big\rangle \delta^{(2)}(z-\om) =-b^2 \xi'(\om)
\label{most}
\ee
as shown in \eq{C7} of Appendix~\ref{appA},
reproducing \rf{29}.

It is also easy to understand that \eq{29} remains valid also for the four-derivative action~\rf{invamin}
(as well as for yet higher derivative actions) if the exponential is defined as the normal product.
This is because $\vp(z)$ pairs with only one $\vp(\om)$ modulo the renormalization of $\alpha$.
Among three parameters $b$, $q$ and $\alpha$ only two are independent. 
For convenience of the computation we can set $\alpha=1$  to be nonrenormalized 
(like in \cite{DK89}) and concentrate on the renormalization of $b_0^2$ and $q_0=1$ or
set $q=1$ to be nonrenormalized 
(like in \cite{KN93}) and concentrate on the renormalization of $b_0^2$ and $\alpha_0=1$.
I keep below  both $q$ and $\alpha$ arbitrary to verify the computations.

The coefficient in front of the first term on the right-hand side 
of \eq{29} represents the conformal weight
of $\e^{\alpha \vp}$. Since  the conformal weight of $\e^{\alpha\vp}$ equals 1,
we arrive at the equation
\be
1=q \alpha-b^2 \alpha^2
\label{DDK1}
\ee
which is nothing but the first of two DDK equations, expressing  $\alpha$ in terms of  $b^2$. 
We have thus argued it remains unchanged for the higher-derivative action.

\section{Computation of the central charge}

The central charge $c^{(\vp)}$ of $\vp$ can be computed from the variation of $T_{zz}$ under the infinitesimal conformal transformation  generated by \rf{hatdel} as
\be
 \LA \hat \delta_\xi T_{zz} (\om)\RA = \frac{ c^{(\vp)}} {12}\xi'''(\om).
\ee 
The normal ordering in $T_{zz}$ is assumed throughout this Section.

\subsection{Quadratic action}

The energy-momentum tensor for the quadratic action is given by \rf{qTzz} with $\eps=0$.
Under the infinitesimal conformal transformation it transforms as
\bea
\LA \hat\delta_\xi  T_{zz}^{(2)} (\om) \RA&=&\frac1{2b^2}  \int \d^2 z  
\big\langle q^2 \xi'''(z)+
 \xi'(z) \p^2 \vp(z)\vp(\om)  
  +\xi(z) \p^3 \vp(z)\vp(\om) \big\rangle  \delta^{(2)}(z-\om)
\non &= &
 \frac {\xi'''(\om)}{2} \left(\frac{q^2}{b ^2} +\frac13-\frac 16 \right)= 
{\xi'''(\om)}\left( \frac{q^2}{2b ^2}+\frac 1{12}\right) , 
 \label{vp2}
\eea
where we have used \eq{A13}.
Here $1/12$ corresponds to the usual quantum addition $1$ to the central charge.
The right-hand side of \eq{vp2} reproduces the DDK formula for the central charge.

Notice that 
the propagator in~\eq{vp2} is {\it exact}\/ as before in \eq{32}. This is why  $b^2$ cancels.

\subsection{Quartic action}

The computation of
$\delta T_{zz}$, where $T_{zz}$ is given by \eq{qTzz} with  $\e^{\vp}$ 
substituted by the renormalized $\e^{\alpha \vp}$, is a bit lengthy 
but easily doable with Mathematica. Equation~\rf{vp2} remains unchanged 
for the quadratic part of $T_{zz}$ which is nonvanishing at $\eps=0$. The variation of the part
which is $\propto\!\eps$ reads for $q=1$ and $\alpha=1$~\cite{Mak23b}
\bea
\lefteqn{\LA \hat\delta_\xi T_{zz}^{(4)} (\om)\RA _{G=0}
= 2 \frac1{b^2} \int \d^2 z\Big\langle \e^{-\vp(\om)} 
\big\{-\xi'''(z) \p\bp\vp (z) +\xi''(z) \p\vp(z)\p\bp \vp(\om)} \non &&
+\xi'(z)\big[2\p^3\bp \vp(z)-\p^2\vp(z)\p\bp\vp(\om)-
\p\bp \vp(z)(\p\vp(\om))^2-\p\vp(z)\p\vp(\om)\p\bp\vp(\om)\big]\non &&
+\xi(z)\big[\p^4\bp \vp(z)-\p^3\vp(z)\p\bp\vp(\om)-
\p^2\bp \vp(z)(\p\vp(\om))^2+\p\vp(z)\p^2\vp(\om)\p\bp\vp(\om)\big] \big\}
\Big\rangle \delta_\eps^{(2)}(z-\om).\non &&
\label{ppp}
\eea

The consideration of the right-hand side of \eq{ppp}
is similar to  \eq{32}  if $T_{zz}$ is defined as the normal product.
Now $\p\vp(z)$ pairs again with only one $\vp(\om)$ modulo the renormalization of the
quadratic in $\vp$ part of $T_{zz}$ which is accounted for in \eq{qTzz} by the change
$b_0^2\to b^2$.
One more $\vp(\om)$ is annihilated by the variational derivative $\delta/\delta \vp(z)$.
We thus obtain 
\bea
\lefteqn{\LA \hat\delta_\xi T_{zz}^{(4)} (\om)\RA_{G=0}
=\frac1{b^2} \int \d^2 z
 \LA 
\big[2 q\alpha \eps \xi'''(z) \p\bp\vp(z)
+(4q\alpha- 2)\eps \xi''(z) \p^2\bp \vp(z) -6\eps \xi'(z) \p^3\bp \vp(z)
 \right. }\non &&
 \left. 
 -4\eps \xi(z) \p^4\bp \vp(z)\big]\vp(\om) \RA
\delta_\eps^{(2)}(z-\om)  
=\frac {\xi'''(\om) }4 \left( -2 \cdot2 q\alpha+(4q\alpha- 2)\cdot 1+6 \frac 23 -4 \frac 12\right) =0,
~~~
\label{vp4}
\eea
where  \eq{Jnm} is used. The propagator   $\LA \vp(z) \vp(\om) \RA$ here 
is exact resulting in the cancellation of $1/b^2$.
Thus the central charge of $\vp$ coincides at $G=0$
with the one for the quadratic action.

We can repeat the computation for the part of $T_{zz}$ in \eq{qTzz} which involves $G$.
Again only the average of the quadratic in $\vp$ part of $\delta T_{zz}$ is nonvanishing 
for the normal ordered $T_{zz}$, 
like we have passed from \rf{ppp} to \rf{vp4}.
This makes the computations pretty much similar to those~\cite{Mak23b} at one loop.
However, 
the contribution of higher loops is effectively  taken into account by the deviations of
 $b^2$, $q$ and $\alpha$ from their bare values. This is why I call it ``intelligent'' one loop.

For the polynomial in derivatives terms we obtain
\bea
\lefteqn{\Big\langle \hat \delta_\xi \frac{1}{b^2}G\eps \big[ (\p\vp)^2\e^{-\alpha\vp} \p\bp\vp-
\p\vp \p(\e^{-\alpha\vp}\p \vp \bp \vp)
+q \p^2(\e^{-\alpha\vp}\p \vp \bp \vp)
-\p(\p\vp \e^{-\alpha\vp} \p\bp \vp)\big]\Big\rangle } \non&&= 
\frac{1}{b^2}Gq\eps \int \d^2 z
\LA \big[-\xi'''(z)\p\bp\vp(z)
-2 \xi''(z) \p^2\bp \vp(z) \big]  
 \vp(\om)
\RA\delta_\eps^{(2)}(z-\om) = 0,~~~~~
\label{tttp}
\eea
where \eq{Jnm} is used.

For the nonlocal term in \rf{qTzz} we analogously find 
\bea
\LA \hat \delta_\xi \left(-\frac{1}{b^2}Gq\eps 
\frac 1\bp \p^2 (\bp\vp \e^{-\vp} \p\bp \vp)\right)\RA&=&
-\frac{1}{b^2}Gq\eps \int \d^2 z\, \xi'''(z)
 \LA\p\bp\vp(z)\vp(\om) \RA 
\delta_\eps^{(2)}(z-\om) \nonumber \\  &=& \frac 12 Gq \xi'''(\om) .
\label{tttnlp}
\eea
In contrast to the  classical case, now a nonvanishing finite contribution arises.
Its occurrence is like in the CFT technique where it has arisen~\cite{Mak22} 
at one loop from the nonlocal (last) term in \rf{Tzz}.

The sum of \rf{vp2}, \rf{vp4}, \rf{tttp} and \rf{tttnlp} determines the central charge of $\vp$.
We thus arrive at the second equation
\be
\frac 6{b^2} +1 +6G q =\frac 6{b^2_0} 
\label{DDK2}
\ee
representing the vanishing of the total central charge. It reproduces DDK for $G=0$.

\section{Change of KPZ-DDK for the four-derivative action}

It is easy to solve Eqs.~\rf{DDK1}, \rf{DDK2} which modify KPZ-DDK for the four-derivative
action \rf{invamin}. The result 
reads
\begin{subequations}
\bea
\tilde b^2 \equiv \alpha^2 b^2 &=& 
\frac {13-d-6\tilde G-\sqrt{(d-d_+)(d-d_-)}}{12(1+\tilde G)}, ~~~~~~
\label{tb}\\
\tilde q \equiv \alpha q &=&1+\tilde b^2,
\label{tq}\\
d_\pm&=&13-6\tilde G\pm 12 \sqrt{1+\tilde G} ,
\label{dmp}
\eea
\label{tbq}
\end{subequations}
\!where $d=26-6/b_0^2$ to comply with the Liouville action.
In \eq{tbq} we have introduced $\tilde G=G/\alpha$, $\tilde b=\alpha b$,  $\tilde q =\alpha q$
which do not depend on the normalization of $\vp$.

For the solution~\rf{tbq} the KPZ barriers of the Liouville theory are shifted to
$d_\pm$ given by \rf{dmp}.
The string susceptibility equals
\be
\gamma_{\rm str}=(h-1) \frac {\tilde q}{\tilde b^2}+2 =
(h-1)\frac{25-d-6 \tilde G+\sqrt{(d-d_+)(d-d_-)}}{12}+2 
\label{sus}
\ee
which extends \eq{gstrfin} to all loops.
It is real for $d>d_+$ where $d_+<25$ for $\tilde G>0$ 
as required for stability of the action~\rf{invamin}.
For example it is real for all $d>1$ if $\tilde G=8$ (then $d_-=-71$).

\section{The relation to minimal models}

As we have seen already in Sect.~\ref{s:dim} the conformal weight of $\e^{\alpha \vp}$ 
shown in \eq{29} coincides with the one for the quadratic action. This occurs because the
terms $\sim\!\eps$ do not revive in the operator product $T_{zz}(z)\e^{\alpha \vp(0)}$
as $\eps\to0$ (in contrasts to   $T_{zz}(z)T_{zz}(0)$). Therefore $\e^{(1-n)\alpha \vp/2}$
are the BPZ null-vectors~\cite{BPZ} as
reviewed in~\cite{ZZ}, Sect.\ 23 for the Liouville action.
 Their conformal weights 
\be
\Delta_{(1-n)/2}=\frac{1-n}2+\frac{1-n^2}4 \tilde b^2
\ee
coincide with the dimensions $\Delta_{1,n}(c)$ of Kac's spectrum of the minimal
models
 \be
\Delta_{m,n}(c)=\frac{c-1}{24} +\frac 14 \left( (m+n)\sqrt{\frac {1-c}{24}}+
 (m-n)\sqrt{\frac {25-c}{24} } \right)^2
\label{Kac}
 \ee
provided that the central charge $c$ is equal to
\be
c=26-d-\frac{\tilde G\big(25-d+6 \tilde G-\sqrt{(d-d_+)(d-d_-)}\big)}{2(1+\tilde G)}.
\label{ctG2}
\ee
It reproduces $c=26-d$ for the Liouville theory as $\tilde G\to0$.

The domain of physically acceptable parameters of the four-derivative action~\rf{invamin} with
$\tilde G>0$, as required by stability of the action, is $d>d_+$ (recall that $d_+=1$ for 
$\tilde G=8$).  It is mapped onto
$c<1$ (and vice versa $d<d_-$ is mapped onto $c>26$)
thus showing of how the KPZ barrier~\cite{KPZ} is bypassed for the four-derivative 
action~\rf{invamin}.

\section{Discussion}

The constant $G$ in the four-derivative action \rf{invamin} is calculable for the Nambu-Goto string.
However, I refrain from saying that the obtained shift of the KPZ barrier is the one for
the Nambu-Goto string. For the Polyakov string, where the part of the emergent
action with four derivatives is $\eps R^2$, the solution for 
$\eps\to0$ is the same as for the Liouville action. Moreover, it has been
 argued~\cite{Mak21,Mak23b} that yet higher-derivative terms 
 of the type   $\eps^{k+1}R\Delta^k R$ with $k\geq 1$ do not
 change this result (the universality takes place). For the Nambu-Goto string there are additional
 higher-derivative terms of the type of the $G$-term in \rf{invamin} which may also
 contribute to the shift. I do not have arguments in favor of the universality in this case.
 
 But the computation of the central charge can be repeated  directly
 for the Nambu-Goto string, 
 using the action ${\cal S}[\vp,\lambda^{ab}]$ that emerges after the path integration 
 over all fields but
 the metric tensor and the Lagrange multiplier $\lambda^{ab}$ which is 
 introduced to have an independent metric tensor.
 The $G$-term in \rf{invamin} and analogous higher-derivative terms emerge from 
 ${\cal S}[\vp,\lambda^{ab}]$
 after the path integration over $\lambda^{ab}$.
This study of the Nambu-Goto string is number one on my list todo!

\subsection*{Acknowledgement}

I am grateful to Alexei Morozov for sharing his insight into string anomalies. 
This work was supported by the Russian Science Foundation (Grant No.20-12-00195).

\appendix

\section{List of formulas for the singular products\label{appA}}

The simplest singular product
\be
\int \d^2 z\, \xi(z) \LA \partial^n \vp(z) \vp(0) \RA 
\delta^{(2)}(z) =(-1)^{n}  \frac{2}{n(n+1)} \xi^{(n)}(0) ,
\label{C7}
\ee
emerges already in a free CFT with the propagator
\be
\LA \vp(z) \vp(0) \RA =8\pi G_0(z), \qquad
G_0(z) =-\frac1{2\pi} \log\Big(\sqrt{z\bz} \mu\Big),
\label{G0}
\ee
where $\mu$ is an infrared cutoff. Equation \rf{C7} can be derived by the formulas
\be
\delta^{(2)}(z) =\bp \frac {1}{\pi z},\qquad
\frac 1{z^n} \bp \frac 1z=(-1)^n \frac1{(n+1)!} \p^n \bp \frac 1z.
\label{B2}
\ee

Equation~\rf{C7} can be alternatively derived introducing 
 the regularization by $\eps$ via the fourth derivative in the action  that results
 in the momentum-space propagator
\be
G_{\eps}(k)=\frac{1}{k^2(1+\eps k^2)},\qquad 
\delta^{(2)}_\eps(k)=\frac{1}{(1+\eps k^2)}.
\label{Geps}
\ee
We then have
\be
8\pi
\int \d^2 z f(z) \p^{n} G_\eps(z-\om)
\delta^{(2)}_\eps(z-\om)= (-1)^n \frac 2{n(n+1)}
\p^n f(\om),
\label{A13}
\ee
reproducing \eq{C7}, and
\be
8\pi 
\int \d^2 z f(z)\big[ -4 \eps\p^{n+1} \bp G_\eps(z-\om)\big]
\delta^{(2)}_\eps(z-\om) = (-1)^n \frac2{(n+1)}
\p^n f(\om).
\label{Jnm}
\ee


\end{document}